\documentclass[prl,showpacs,superscriptaddress,twocolumn,preprintnumbers,amssymb,amsfonts]{revtex4}
\usepackage{graphicx}
\usepackage{amssymb}
\usepackage{amsmath}
\usepackage{bm}

\begin{document}

\input epsf
\def\beq{\begin{equation}}
\def\eeq{\end{equation}}
\def\beqn{\begin{eqnarray}}
\def\eeqn{\end{eqnarray}}
\def\etal{\emph{et al.}}
\renewcommand{\v}[1]{\boldsymbol{#1}}
\newcommand{\up}{\uparrow}
\newcommand{\down}{\downarrow} 

\title{Topological surface states in three-dimensional magnetic insulators}
\author{Joel~E.~Moore}
\affiliation{Department of Physics, University of California,
Berkeley, CA 94720} \affiliation{Materials Sciences Division,
Lawrence Berkeley National Laboratory, Berkeley, CA 94720}
\author{Ying Ran}
\affiliation{Department of Physics, University of California,
Berkeley, CA 94720} 
\author{Xiao-Gang Wen}
\affiliation{Department of Physics, Massachusetts Institute of Technology,
Cambridge, MA 02115}
\date{\today}

\begin{abstract}
An electron moving in a magnetically ordered background feels an effective magnetic field that can be both stronger and more rapidly varying than typical externally applied fields.  One consequence is that insulating magnetic materials in three dimensions can have topologically nontrivial properties of the effective band structure.  For the simplest case of two bands, these ``Hopf insulators'' are characterized by a topological invariant as in quantum Hall states and $\mathbb{Z}_2$ topological insulators, but instead of a Chern number or parity, the underlying invariant is the Hopf invariant that classifies maps from the 3-sphere to the 2-sphere.  This paper gives an efficient algorithm to compute whether a given magnetic band structure has nontrivial Hopf invariant, a double-exchange-like tight-binding model that realizes the nontrivial case, and a numerical study of the surface states of this model.
\end{abstract}
\maketitle


Recent theoretical and experimental work has shown that there exist nonmagnetic band insulators in which spin-orbit coupling plays a role similar to that of the magnetic field in the integer quantum Hall effect (IQHE).  In two dimensions~\cite{km2}, these ``topological insulators'' have robust edge states, observed in HgTe/(Hg,Cd)Te heterostructures~\cite{molenkampscience}, and are predicted to show a spin quantum Hall effect.  The existence of a genuinely three-dimensional topological insulator phase~\cite{fu&kane&mele-2007,moore&balents-2006,rroy3D} with protected {\it surface} states,
recently observed in Bi$_{0.9}$Sb$_{0.1}$~\cite{hsieh},
is rather surprising because the IQHE does not have a fully three-dimensional version, but only layered versions of the 2D case.  Both 2D and 3D topological insulators are nonmagnetic, and in fact unbroken time-reversal invariance is required for the edge state to remain gapless.  The edge or surface states of topological insulators and IQHE states exist because there are topological invariants that distinguish these insulating states from ordinary insulators, and across a boundary between one of these states and an ordinary insulator, the energy gap must close.

The goal of this paper is to show that there are genuinely three-dimensional (i.e., not layered)
topological insulating phases of electrons moving in a magnetic background.
The problem of electrons moving in such a background has attracted
considerable interest because of its relevance to materials in which
electrons from outer orbitals move in a magnetic environment generated by the
ordered magnetic moments of core electrons.  The outer-orbital electrons are
frequently described by tight-binding models in which the
hopping of an electron from site $i$ to site $j$ depends on its initial and
final spin and the nearby core spin configuration.  Several materials with layered kagome structures derived from a parent pyrochlore lattice, such as Nd$_2$Mo$_2$O$_7$~\cite{taguchi}, have been argued via this mechanism to show the quantum anomalous Hall effect~\cite{ohgushi,kagomeahe}, in which electrons hopping in a magnetic background form a two-dimensional integer quantum Hall state.  We give an explicit cubic-lattice model with a nontrivial {\it three-dimensional} topological invariant (the
Hopf invariant in momentum space) and extended surface states and discuss in
closing which materials might realize the ``Hopf insulator'' phase.  This
insulator is simpler in some ways than the $\mathbb{Z}_2$ topological
insulators, since the minimal realization requires only two bands (counting
spin) rather than four~\cite{moore&balents-2006}.  We argue that pyrochlore-lattice compounds with noncollinear magnetic order are realistic candidates for Hopf insulators, and discuss specific materials in closing.

First, let us give a general introduction to topological band insulators in any
dimension. Consider a band insulator in $d$ dimensions with $n$ filled bands and
$m$ empty bands. In ${\v k}$ space, such a band insulator is described by
$m+n$ dimensional matrix $H({\v k})$ which has $m$ positive and $n$ negative
eigenvalues for any ${\v k}$ (assuming $E_F=0$).  Without changing the
ground state, we may deform all
the positive eigenvalues to $1$ and all the positive eigenvalues to $-1$.  Thus $H({\v k})$ has the form
\begin{equation}
 H({\v k})=W({\v k})I_{m,n}W^\dagger({\v k}),
\end{equation}
where $I_{m,n}$ is the diagonal matrix with $m$ 1's and $n$ $-1$'s on the diagonal,
and $W({\v k}) \in SU(m+n)$.  We see that for any fixed ${\v k}$, $H({\bf k})$
is a point on the manifiold $SU(m+n)/G_{m,n}$, where $G_{m,n}$ is a  subgroup
of  $SU(m+n)$ that is formed by the transformations that leave $I_{m,n}$
invariant. We find that $G_{m,n}=SU(m)\times SU(n)\times U(1)$ and
\begin{equation*}
H({\v k}) \in SU(m+n)/SU(m)\times SU(n)\times U(1) \equiv M_{m,n}.
\end{equation*}
So a band insulator is described by a mapping from the Brillouin
zone $T^d$ (a $d$-dimensional torus) to $M_{m,n}$.  The different classes of
topological band insulators in $d$ dimensions are just classes of
mappings $T^d \to M_{m,n}$.  Time-reversal invariance in fermionic systems (or other additional symmetries~\cite{schnyder}) imposes additional restrictions that can separate topological
classes.

For two-band insulators, we have $M_{1,1}=S^2$. Mappings from $S^3\to S^2$
can be non-trivial (since the third homotopy group $\pi_3(S^2)=\mathbb{Z}$), and a familiar example to physicists is the Hopf map:
\begin{equation*}
\chi \to {\v n}=\chi^\dagger {\v \sigma } \chi ,
\end{equation*} 
where $\chi^T=(z_\up,z_\down)$ and $|z_\up|^2+|z_\down|^2=1$.
The non-trivial Hopf map $S^3\to S^2$ implies that there can be non-trivial
mappings $T^3\to S^2$. The corresponding two-band topological insulators in 3D will be called Hopf insulators; these insulators must break time-reversal in order to avoid Kramers degeneracies at time-reversal-invariant points ${\bf k} = -{\bf k}$ in the Brillouin zone.  Before writing down a specific model of a Hopf insulator, we review briefly the basic ideas of Chern number and the Hopf invariant and give an algorithm to compute the Hopf invariant for any two-band model.

There are two equivalent ways to understand the Chern number of a gapped 2D band structure with two bands, and both are helpful in understanding the Hopf invariant in 3D.  Such a band structure is given by four real, periodic functions of ${\v k}$: the Bloch Hamiltonians are
\beq
H({\v k}) = a_1({\v k}) \sigma_x + a_2({\v k}) \sigma_y + a_3({\v k}) \sigma_z + a_4({\v k}) {\v 1}.
\label{components}
\eeq
The fourth function $a_4({\v k})$ plays no role in the topological classification and is omitted in the following, since it only shifts the energy levels, but can be experimentally important in determining if a material is insulating.  The gapless condition is $a_1({\v k})^2 + a_2({\v k})^2 + a_3({\v k})^2 > 0$.  Let the components of ${\v k}$ run from $- \pi$ to $\pi$.  For each value of ${\v k}$, a direction on the unit sphere is fixed by
\beq
{\hat {\v n}}({\v k}) = {(a_1({\v k}),a_2({\v k}),a_3({\v k})) \over \sqrt{a_1({\v k})^2 + a_2({\v k})^2 + a_3({\v k})^2}}.
\eeq
Now in two dimensions, the Brillouin zone has the topology of the torus $T^2$, and to classify band structures we need to classify maps from $T^2$ to $S^2$.  Because any map from the circle to the sphere $S^2$ can be contracted to a point, maps from the torus to $S^2$ are equivalent topologically to maps from $S^2$ to $S^2$~\cite{ass}. 
These maps are classified by an integer ``homotopy invariant'' that generalizes the notion of winding number; it counts how many times the first sphere wraps the second.  Going back to the torus, the invariant is
\beq
n=\int_{-\pi}^\pi\,dk_x\,\int_{-\pi}^\pi\,dk_y\,j_z,
\eeq
where the local current (to be used later in 3D) is
\beq
j_{\mu} = {1 \over 8 \pi} \epsilon_{\mu \nu \lambda} {\hat{\v n}} \cdot (\partial_\nu {\hat{\v
n}} \times \partial_\lambda {\hat{\v n}}).
\label{currdef}
\eeq
The same invariant may be more familiar in terms of the ground state spinor $|\chi(k_x,k_y)\rangle$.  While this spinor has a $U(1)$ ambiguity, the ``Berry flux''
\beq
F = {i \over 2 \pi} \left[ \left\langle {\partial \chi \over \partial k_x} \Big| {\partial \chi \over \partial k_y} \right\rangle - \left\langle {\partial \chi \over \partial k_y} \Big| {\partial \chi \over \partial k_x} \right\rangle \right]
\eeq
is gauge-invariant and equal to $j_z$.  For the three-dimensional case, we need to use the classification of maps from $T^3$ to $S^2$ found by Pontryagin~\cite{pontryagin}.


The classification of two-band band structures in 3D, i.e., maps from $T^3$ to $S^2$, is complicated because $T^3$ includes three 2D tori, and each of these may have a nontrivial Chern number.  The guess that such a band structure is classified by three Chern numbers, which if nonzero give layered integer quantum Hall states, plus the Hopf invariant is actually incorrect, and in fact maps from $T^3$ to $S^2$ do not form a group.  Here we will focus on the case where the Chern numbers vanish and there is no quantum Hall effect.  Otherwise the Hopf invariant is no longer integer-valued but takes values in the finite group $\mathbb{Z}_{2 \cdot {\rm GCD}(n_x,n_y,n_z)}$ for Chern numbers $n_i$~\cite{pontryagin,misirpashaev}.  A geometrical picture of the Hopf invariant is obtained by noting that each point on $S^2$ has a preimage that is a circle in $T^3$, and that the linking number of two such circles taken from different points of $S^2$ is just the Hopf invariant.  For evaluating the invariant, it is easier to use the integral, following Wilczek and Zee~\cite{wilczekzee}\footnote{It appears that there is an erroneous factor of $1/(2 \pi)$ in equation of Ref.~\onlinecite{wilczekzee}.},
\beq
n_h = -\int d^3k\, {\v j} \cdot {\v A}.
\label{hopfdef}
\eeq
Here the gauge field ${\v A}$ satisfies the magnetostatic equation $\nabla \times {\v A} = {\v j}$.  This integral is invariant under a small variation $\delta{\hat{\v n}}$ in (\ref{currdef}) and hence a homotopy invariant.

The Hopf invariant is similar to the $\mathbb{Z}_2$ invariant or Chern parity~\cite{fu&kane1-2006,essinmoore} in topological insulators in that its standard integral expression uses the gauge-dependent quantity ${\v A}$ even though the final result (\ref{hopfdef}) is gauge-invariant when the Chern numbers are zero; alternately, one can use an explicitly gauge-invariant but nonlocal form~\cite{jackiwrev}.  The Abelian Chern-Simons form that appears in the integrand gives an invariant that is fully gauge-invariant (again, when all Chern numbers are zero).  One previous appearance of the Hopf invariant in physics is in determining the statistics of solitons in the nonlinear $\sigma$ model~\cite{wilczekzee}; as an example, a magnetic moment field ${\bf \hat n}({\bf r})$ coupled to fermions can have the statistics of its solitons changed by the Hopf term generated by integrating out ``fast'' fermions coupled to the field~\cite{hsianglee}.

\begin{figure}[!ht]
\scalebox{0.75}{\includegraphics{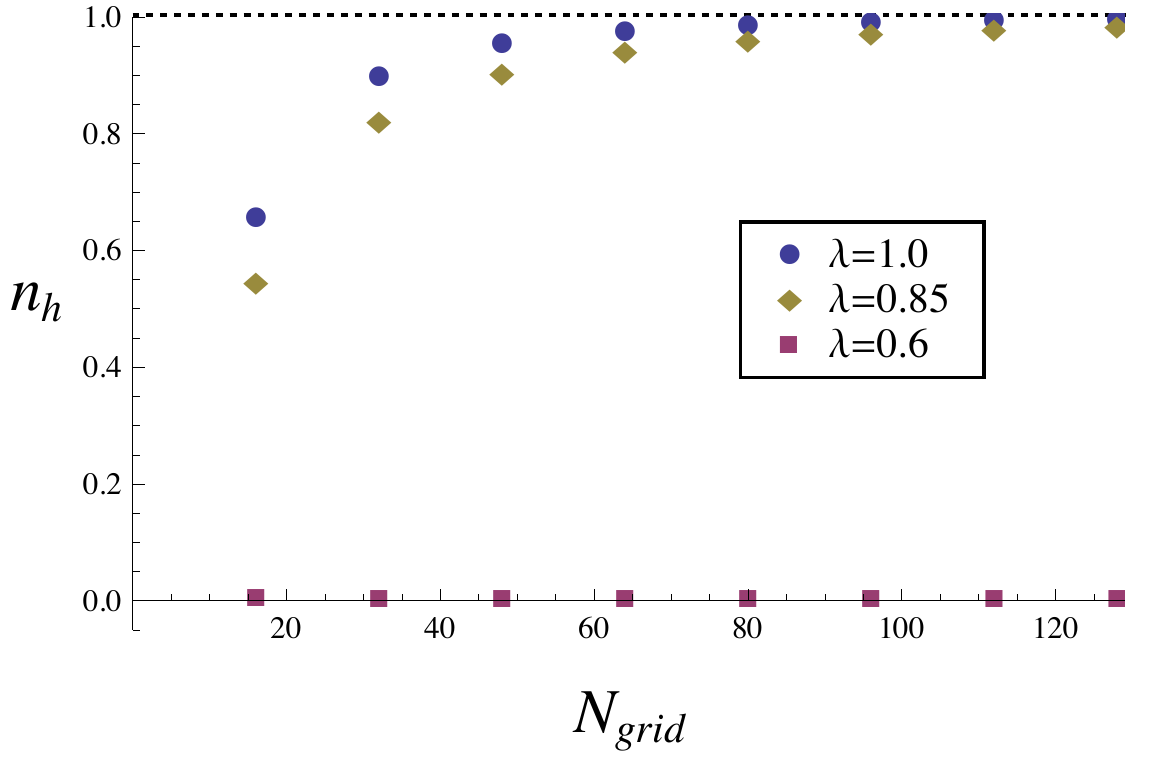}}
\caption{Convergence of the numerical Hopf invariant integral (\ref{hopfdef}) to 0 or 1 as number of grid points increases.  The parameter $\lambda$ is the reduction of the hopping elements relative to the on-site term in the model (\ref{theham}).  The horizontal axis is the linear size of the grid ($N_{tot} = N_{grid}^3$).}
\end{figure}

The Hopf invariant can be computed rapidly for any two-band problem with zero Chern numbers.
We compute the Hopf invariant numerically by discretizing the torus in real space to compute ${\v j}$, then Fourier transforming and solving the magnetostatic equation in momentum space in the discretized version of the gauge $\partial_\mu A_\mu = 0$.  This results in an $O(N \log N)$ algorithm for $N$ points on the Brillouin zone grid.  We studied a particular local tight-binding model as an example of nontrivial Hopf invariant, and confirmed that it has the same Hopf invariant and vanishing Chern numbers as the Hopf map to $S^2$.  Fig. 1 shows the resulting values of (\ref{hopfdef}) for grids of different sizes for the model
\beqn
z_\uparrow &=& \sin k_x + i \sin k_y, \cr
\quad z_\downarrow &=& \sin k_z + i \left( \cos(k_x)+\cos(k_y)+\cos(k_z)-\frac{3}{2}\right), \cr
v^i &=& {\v z}^\dagger \sigma^i {\v z},\quad H = {\v v} \cdot {\v \sigma}.
\label{theham}
\eeqn
This model is constructed to be continuously connected to the Hopf map, but with only low Fourier components and hence a local tight-binding representation in real space.  Its Hopf invariant is +1 (Fig. 1) and Chern numbers are zero.  It has an $E \leftrightarrow -E$ symmetry (since $a_4=0$ in (\ref{components})) and no states in the band gap $-0.25 < E < 0.25$.

As a check, the hopping terms in the model (\ref{theham}) can be reduced by a factor $\lambda$ while keeping the on-site $\sigma_z$ term fixed.  This leads to a gapless region for approximately $0.63 < \lambda < 0.82$.  To confirm that this gap closing is associated with a phase transition between the ordinary and topologically nontrivial states, we computed the Hopf invariant for different values of $\lambda$ (Fig. 1).
In the computation of midgap surface states below, we also checked that there are no midgap states for a boundary between two slabs with $\lambda_1$ and $\lambda_2$ either both greater than or both less than this gapless range, while the surface state appears once one is above and one is below.

In real space, there are three terms in the tight-binding Hamiltonian obtained by computing Fourier components of (\ref{theham}): an on-site Zeeman interaction $-13 \sigma_z /4$; a nearest-neighbor spin-dependent hopping of strength $3 \lambda/2$; and length-2 spin-dependent hoppings of strength $\lambda/2$.  These specific hoppings and the symmetry of (\ref{theham}) are not required in order to ensure that the topological invariant takes the value 1, and any change to the model that does not close the gap will not change the topological properties.  The topological invariant guarantees that a {\it smooth} (i.e., adiabatic) boundary between a Hopf insulator such as (\ref{theham}) and vacuum or a trivial insulator will have gapless excitations.

\begin{figure}[!ht]
\scalebox{0.65}{\includegraphics{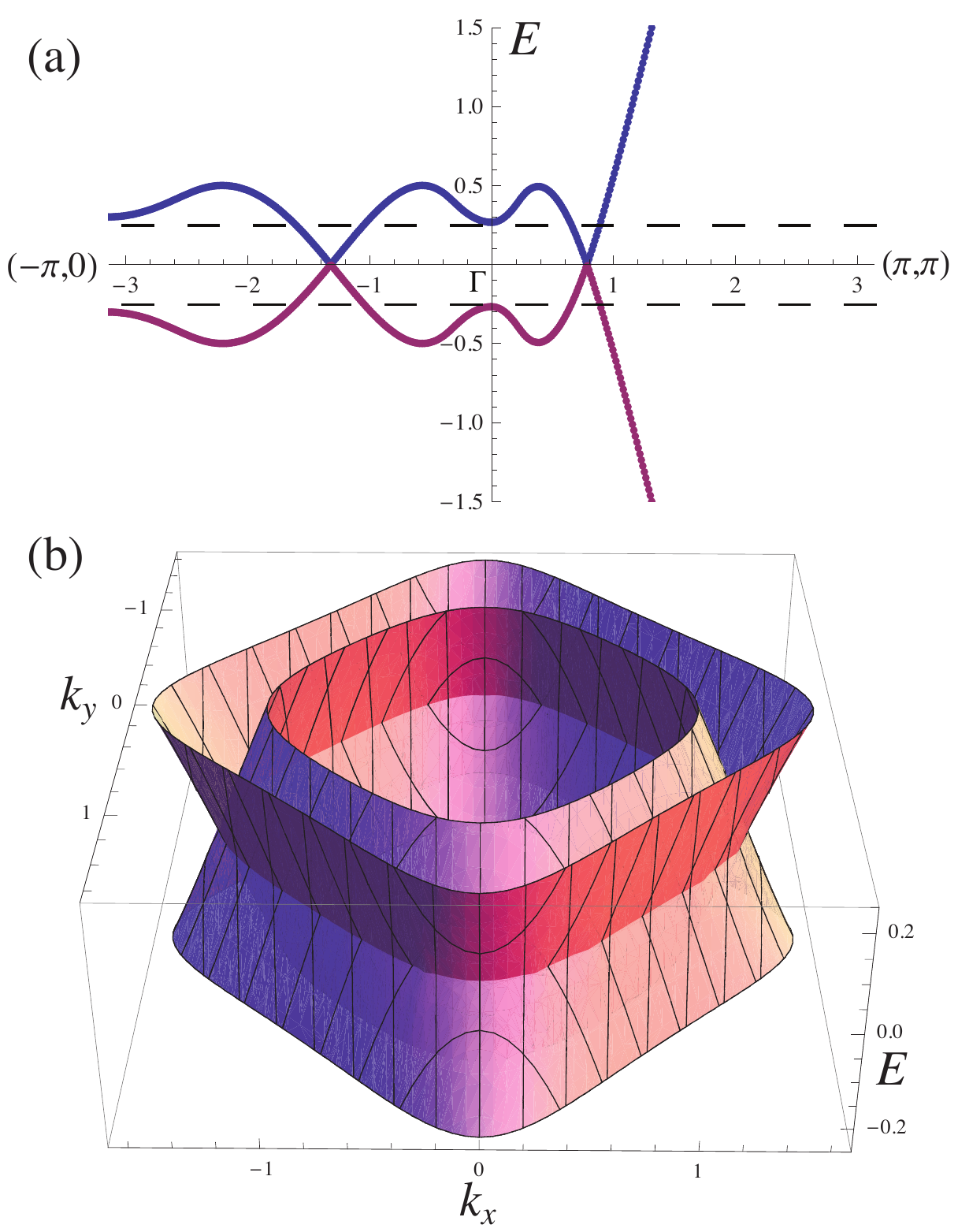}}


\caption{(a) The lowest electron and hole bands with sharp (001) surfaces in the model (\ref{theham}), for a 16-site-thick slab.  The Brillouin zone is $-\pi \leq k_x < \pi$, $-\pi \leq k_y < \pi$.  The bulk band gap is shown by the dotted lines at $E = \pm 0.25$, and the spectrum is cut off at $E = \pm 1.5$. (b) 3D plot $(k_x,k_y,E)$ of midgap dispersion of (001) surface states near $\Gamma$ point.}
\end{figure}

We find numerically that a gapless surface state still exists even for (at least some) sharp boundaries, as in integer quantum Hall and topological insulator states, where a sharp boundary between materials of different topological invariant generates an edge or surface state.  We computed numerically the midgap surface states induced by abrupt boundaries in the (001), (010), and (100) directions.  In all cases we find gapless, dispersive surface states from a single midgap band; there is a single Dirac point except for the (001) direction, which has a ring of gapless points (Fig. 2).  Adding perturbations that break particle-hole symmetry shifts the band-crossing energy by different amounts for different points on this ring.  It is not clear that the surface state survives for all sharp surfaces, but for the surfaces above, a midgap surface state appears precisely when the Hopf invariant changes.


The extension to multiple bands is more subtle mathematically for the Hopf insulator than for the IQHE or $\mathbb{Z}_2$ insulator.   (Of course, extra bands that do not mix with the Hopf-nontrivial bands will not modify consequences such as the surface state.)  In the IQHE, each band carries its own integer invariant, with a zero sum rule over all the bands~\cite{ass}.  In the $\mathbb{Z}_2$ insulator, each Kramers pair carries one $\mathbb{Z}_2$ invariant, again with a zero sum rule~\cite{moore&balents-2006}.  In these cases, the physical state of an electronic system is determined by a sum over the (possibly degenerate) occupied bands.  However, the generalization of the Hopf invariant, the integration of the nonabelian Chern-Simons form, is gauge-dependent once there are degenerate occupied bands: the integral of the nonabelian Chern-Simons form changes by an integer under large gauge transformations.

The multiple-band problem is related to how disorder affects Hopf insulators.  For both Chern number and Chern parity (the $\mathbb{Z}_2$ invariant of a noninteracting disordered system~\cite{essinmoore}), each individual state or Kramers pair of a large disordered sample can be assigned one invariant, and the state is stable until nontrivial states ``float'' across the Fermi level.  If the surface state of the Hopf insulator is stable to disorder, then the surface theory must contain some topological invariant as in the 3D $\mathbb{Z}_2$ insulator's surface~\cite{ryu&alii-2007,ostrovskymirlin}, unlike the 2D case where time-reversal symmetry is sufficient~\cite{xumooreedge,congjun}.  Even if the surface state is localized, as likely since there is no obvious surface topological invariant, it may still be observable by photoemission or tunneling experiments, or in clean samples.

The general requirements for a magnetic material to be in the 3D Hopf insulator phase because of spin-orbit coupling of itinerant electrons to a magnetic background are similar to those required for the 2D quantum anomalous Hall effect, believed to be relevant to the metallic frustrated magnet Nd$_2$Mo$_2$O$_7$~\cite{taguchi,ohgushi,kagomeahe}.  Noncollinear static magnetic order of core electrons occurs in many 3D frustrated materials, including the parent pyrochlore lattice of the kagome lattice that has primarily been studied~\cite{ohgushi,kagomeahe} (the two-dimensional kagome lattice is obtained as one layer from a pyrochlore structure).  Although the pyrochlore lattice, e.g. in ``spin ice'' compounds such as Dy$_2$Ti$_2$O$_7$, may lack long-range magnetic order, many pyrochlore compounds spin-Peierls distort at low temperature into a lower-symmetry phase with noncollinear order.

Hence a realistic microscopic model starts from local Hund's rule coupling of outer-orbital electrons to the magnetic order of a weakly distorted pyrochlore.  Realization of the specific cubic lattice example discussed above seems unlikely in a real solid, although it may be possible to simulate this Hamiltonian using the tunable spin-dependent hoppings of ultracold atoms in optical lattices.  An important future direction is to calculate the Hopf invariant for effective band structures in frustrated magnetic compounds such as R$_2$Mo$_2$O$_7$, R a rare earth ion.  Experimental searches in such compounds for surface states using photoemission or tunneling methods might reveal the Hopf insulator state directly.  Coupling of conduction electrons to complex magnetic order has also been discussed in perovskite manganites~\cite{ye}.




The authors thank A. Abanov, F. Guinea, D.-H. Lee, and A. Vishwanath for helpful
conversations, and NSF DMR-0238760 and DMR-0804413 (JEM), ARO/DARPA (YR) and
NSF DMR-0706078 (XGW) for financial support.


\end{document}